\documentclass{svproc}

\usepackage{url}
\usepackage{amsmath}
\usepackage{graphicx}
\usepackage{array}
\newcommand{\PreserveBackslash}[1]{\let\temp=\\#1\let\\=\temp}
\usepackage{multirow}
\newcolumntype{C}[1]{>{\PreserveBackslash\centering}p{#1}}
\newcolumntype{R}[1]{>{\PreserveBackslash\raggedleft}p{#1}}
\newcolumntype{L}[1]{>{\PreserveBackslash\raggedright}p{#1}}

\usepackage{cite}
\usepackage{amssymb,amsfonts}
\usepackage{algorithmic}
\usepackage{algorithm}
\usepackage{textcomp}
\usepackage{xcolor}
\usepackage{setspace}
\usepackage{pifont}

\begin{document}
\mainmatter              

\title{EchoFlow: A Workload-Aware Parameter Tuning Method for Blockchain Systems}

\titlerunning{EchoFlow: A Workload-Aware Parameter Tuning Method}

\author{Ben Lian$^{1,2}$, Linpeng Jia$^{1}$, Xing Chen$^{1,2}$, Xiaofeng Chen$^{3,4}$, Yi Sun$^{1,2,5,6}$\textsuperscript{(\ding{41})}}

\institute{Institute of Computing Technology, Chinese Academy of Sciences, Beijing 100190 
 \and
School of Computer Science and Technology, UCAS, Beijing 100049 \and
Hangzhou Qulian Technology Co., Ltd., Hangzhou 310051 \and
State Key Laboratory of Blockchain and Data Security, ZJU, Hangzhou 310027 \and
Beijing Advanced Innovation Center for Future Blockchain and Privacy Computing, Beihang University, Beijing 100191 \and 
Shandong Key Laboratory Blockchain Finance, Jinan 250014
\\ \email{sunyi@ict.ac.cn}}

\maketitle

\begin{abstract}  
Blockchain systems expose a large number of tunable parameters that significantly influence system performance. However, in practice, a single parameter configuration is often applied across different workloads, leaving substantial unexploited performance potential. To address this, we propose EchoFlow, a blockchain parameter tuning framework that adaptively adjusts parameter configurations based on workload characteristics, enabling continuous performance optimization. EchoFlow employs a distributed reinforcement learning approach in which multiple actors perform parallel sampling to mitigate the substantial time required for sample generation in blockchain environments. To further accelerate convergence, we introduce a genetic algorithm during the initial phase of training to generate high-quality samples. Extensive experimental evaluations demonstrate that EchoFlow consistently outperforms existing methods across diverse workload scenarios while also reducing training time, highlighting its effectiveness and practical value.

\end{abstract}

\section{Introduction}

\begin{figure}[htbp]
\centering
\begin{minipage}[t]{0.50\linewidth}
\centering
\includegraphics[width=\linewidth]{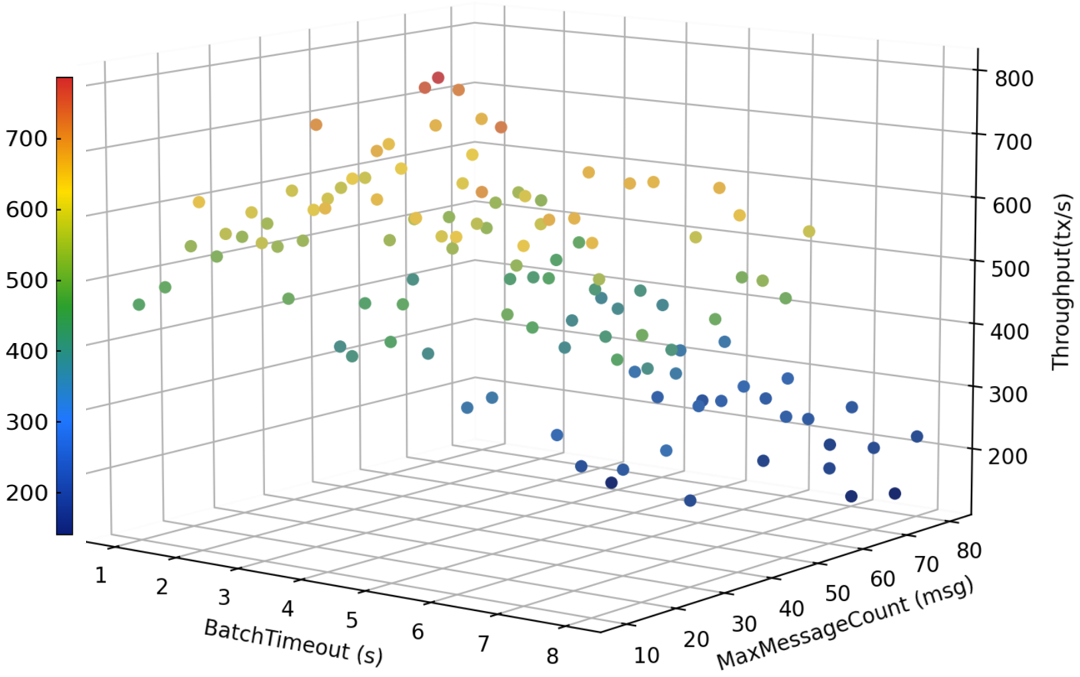}
\caption{\centering Throughput variations under two-parameter tuning}
\label{fig:1_1}
\end{minipage}\hfill
\begin{minipage}[t]{0.47\linewidth}
\centering
\includegraphics[width=\linewidth]{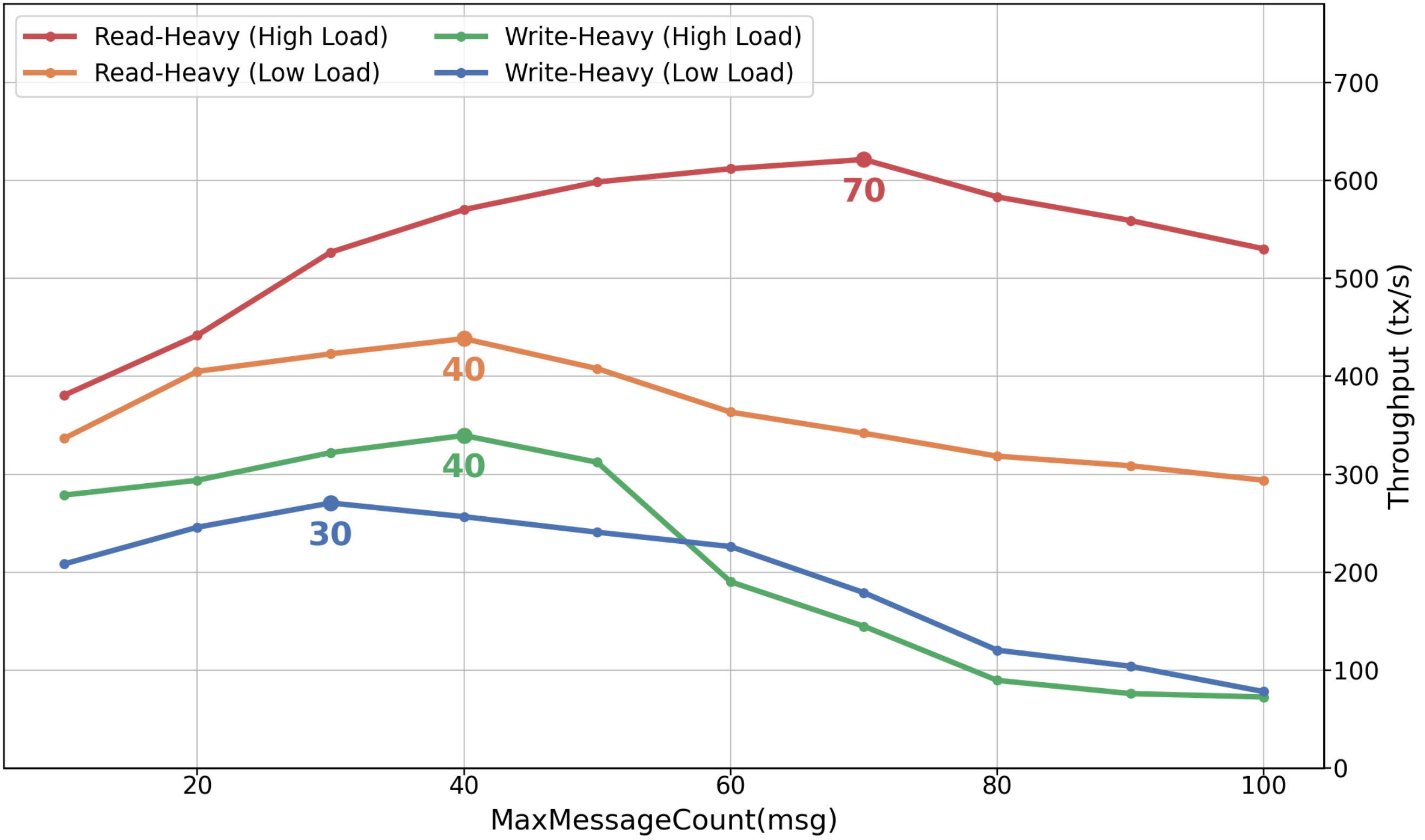}
\caption{\centering Optimal parameter under different workload patterns}
\label{fig:1_2}
\end{minipage}
\end{figure}

As a distributed system, the operation of a blockchain system relies on numerous preconfigured system parameters. These parameters affect all phases of the blockchain system, including peer-to-peer communication, consensus  execution, and transaction execution. Consequently, parameter configuration has a significant impact on system performance. As illustrated in Fig.~\ref{fig:1_1}, adjusting only two parameters in Hyperledger Fabric\cite{hyperledgerFabricDocs} leads to substantial performance variations.

This also implies that appropriate parameter tuning can significantly enhance the performance of blockchain systems. For instance, prior research\cite{sharma2019blurring,xu2023adaptchain} has shown that properly adjusting the number of transactions per block can significantly improve blockchain throughput, and that further optimization of this parameter leads to continued performance gains until the system converges to a stable state. More importantly, the parameter tuning approach does not alter the core operational logic of the blockchain system, indicating that it is independent of existing performance optimization techniques. Given that current blockchain platforms are relatively mature yet still have room for performance improvement, this method enables blockchain systems to approach their theoretical performance limits.

However, determining appropriate parameter configurations is far from trivial. Blockchain systems expose a large number of tunable parameters, and complex nonlinear interdependencies exist among them. Consequently, blockchain parameter tuning constitutes a high-dimensional optimization problem in a multi-parameter space. In practice, parameter configuration in most blockchain deployments is still conducted manually or based on expert-recommended settings. Such approaches are not only inefficient but may also be unable to achieve optimal performance.

To address this issue, various automatic parameter tuning techniques have been proposed, among which deep reinforcement learning (DRL)-based approaches\cite{li2023autotuning,lin2024tunechain} have attracted particular attention and have gradually become mainstream. Compared with traditional machine learning or search-based methods, DRL-based approaches can learn effective parameter adjustment policies through trial-and-error exploration and typically achieve superior performance. Nevertheless, existing DRL-based methods still face several unresolved challenges in real-world blockchain systems.

\textbf{Challenge 1: How to Achieve Workload-Aware Dynamic Parameter Tuning?} In practice, blockchain systems operate under dynamically varying workloads\cite{zhang2024morphdag}, and the optimal parameter configuration is not static across different workload conditions. As illustrated in Fig.~\ref{fig:1_2}, we measure the throughput of Hyperledger Fabric under four different workloads, showing that the optimal parameter setting varies across distinct workload patterns. Therefore, the ability to dynamically adjust system parameters in response to workload changes is critical to the effectiveness of optimization methods in practical deployments. However, existing studies do not fully support dynamic parameter adaptation based on workload variations. For example, Athena\cite{li2023autotuning} primarily trains its optimization model under a single workload scenario, while TuneChain\cite{lin2024tunechain} trains its model using the transaction conflict ratio as the workload characteristic, which may not capture finer-grained workload variations.

\textbf{Challenge 2: How to Accelerate DRL Training Under Diverse Workloads?} The training of DRL models relies on interactions between the agent and the environment. However, a critical bottleneck arises during interaction with the blockchain system: each parameter adjustment requires a system restart for the new configuration to take effect. This process incurs substantial overhead, making performance feedback costly and significantly limiting training efficiency, especially when large numbers of samples are required under diverse workloads.

In this paper, we design and implement a workload-aware automatic parameter tuning approach, EchoFlow, for blockchain systems. Compared with existing DRL-based approaches\cite{li2023autotuning,lin2024tunechain} for blockchain parameter tuning, our work further addresses the two aforementioned critical challenges.

\textbf{To tackle Challenge 1}, we analyze blockchain workload characteristics and model them as a workload feature vector that captures three dimensions: workload scale, transaction type, and access pattern. We then extend the original DRL framework by incorporating this workload feature vector into the state space, enabling the trained DRL agent to dynamically adjust parameter configurations according to varying workload types. This design facilitates adaptive performance optimization under dynamic workload conditions.

\textbf{To tackle Challenge 2}, we propose a distributed DRL training method, in which multiple distributed-actors interact with the environment in parallel to collect samples. 
Furthermore, we introduce a genetic algorithm-based method to generate a large number of high-quality samples covering representative workload scenarios.
These designs mitigate the substantial time required for sample generation in blockchain environments, thereby accelerating training.

We summarize our main contributions as follows:

\begin{itemize}
\item[$\bullet$] We model blockchain workload characteristics and propose EchoFlow, a workload-aware parameter tuning framework  that adaptively adjusts system configurations according to observed workload patterns, enabling continuous performance optimization.
\item[$\bullet$] We employ a distributed DRL training method with multiple actors performing parallel sampling to reduce the time required for sample generation, and introduce a genetic algorithm-based method to generate high-quality initial samples for DRL training, improving training efficiency.
\item[$\bullet$] Extensive experimental evaluations show that, compared with existing approaches, our method dynamically adjusts parameters under diverse workload conditions, achieves superior performance, and effectively accelerates the training process.
\end{itemize}

\section{Formulation}

\subsection{Blockchain Parameters} 

The Execute-Order-Validate (EOV) architecture is a representative blockchain architecture aimed at improving performance and flexibility. It is widely adopted in permissioned blockchain scenarios, with Fabric\cite{hyperledgerFabricDocs} serving as a prominent implementation. 
Using Fabric as an example, we analyze in detail how numerous parameters in a blockchain system influence different stages.
\vspace{-5mm}
\begin{table}[H]
\centering
\caption{Tunable parameters in Fabric}
\label{tab:eov_params}
\begin{tabular}{p{2.8cm}p{7.6cm}}
\hline
\textbf{Phase} & \textbf{Tunable Parameters} \\
\hline
Execution Phase & \texttt{CORE\_PEER\_CLIENT\_CONNTIMEOUT}, \texttt{CORE\_PEER\_AUTHENTICATION\_TIMEWINDOW}, \texttt{CORE\_PEER\_KEEPALIVE\_CLIENT\_INTERVAL}, ... \\

Ordering Phase & \texttt{ORDERER\_RAMLEDGER\_HISTORYSIZE}, \texttt{MaxMessageCount}, \\ & \texttt{PreferredMaxBytes}, ... \\

Validation Phase & \texttt{CORE\_PEER\_GOSSIP\_PULLINTERVAL}, \texttt{CORE\_PEER\_GOSSIP\_SENDBUFFSIZE},                            \texttt{CORE\_PEER\_GOSSIP\_STATE\_BATCHSIZE}, ... \\
\hline
\end{tabular}
\end{table}
\vspace{-5mm}

\textbf{Execution Phase.} The client sends transaction proposals to peer nodes, and each peer verifies and executes the proposal. The execution results are not immediately committed to the local ledger; instead, they are returned to the client as signed proposal responses. The tunable parameters in this phase primarily affect the communication between clients and peer nodes.

\textbf{Ordering Phase.} After collecting a sufficient number of proposal responses, the client packages them and submits them to the ordering nodes. The leader ordering node sorts all transactions according to the consensus protocol (e.g., Raft) and packages them into blocks. The tunable parameters in this phase mainly affect message reception by ordering nodes and block formation.

\textbf{Validation Phase.} The ordering nodes broadcast newly generated blocks to peer nodes, and these blocks are also synchronized across peers. Upon receiving a block, each peer validates the transactions contained within it and immutably commits valid transactions to its local ledger. The tunable parameters in this phase mainly affect block dissemination and synchronization.

\subsection{Workload Characterization}

Characterizing blockchain workloads requires a systematic analysis to capture the complexity of real-world deployments. In this work, we summarize blockchain workloads as three core characteristics---\textbf{workload scale}, \textbf{transaction type}, and \textbf{access pattern}---to facilitate effective differentiation among heterogeneous workload scenarios.

\textbf{1) Workload Scale.}
In real blockchain networks, request loads are highly dynamic and often non-stationary, frequently exhibiting unpredictable and irregular burst patterns\cite{zhang2024morphdag}. For example, in Ethereum, the instantaneous transaction request rate---measured in queries per second (QPS)---can reach up to 6.1 times the daily minimum value. To quantitatively capture such dynamics, we use two variables, \(\mu\) and \(\beta\), to represent the QPS and the degree of QPS variation. They are defined as follows:
\(\mu = \mathbb{E}[\lambda(t)]\), \(\beta = {\sqrt{\mathbb{E}[(\lambda(t) - \mu)^2]}}/{\mu}\), where \(\lambda(t)\) denotes the instantaneous QPS. A larger \(\beta\) indicates stronger burstiness, and \(\beta > 1\) implies substantial workload variability.

\textbf{2) Transaction Type.}
Transactions in blockchain systems are heterogeneous, and different transaction types exhibit distinct performance bottlenecks\cite{khan2025scalability}. For instance, read operations are primarily limited by state lookup efficiency, whereas write operations involve state updates, are more affected by the consensus process, and may introduce conflicts when accessing shared data. Therefore, it is necessary to characterize both read and write operations within transactions. 

Accordingly, we define each transaction \(\tau_i\) as \((r_i, w_i)\), where \(r_i\) denotes the read intensity, representing the number of state data retrievals, and \(w_i\) denotes the write intensity, representing the number of state modification operations. For example, a typical transfer transaction can be defined as \(\tau_t = (2, 2)\).

\textbf{3) Access Pattern.}
Access to blockchain state data is typically non-uniform and exhibits significant skewness. A small number of hot accounts are accessed frequently, whereas the majority of accounts receive minimal access\cite{zhang2024morphdag}. Therefore, we model the access probability of the \(k\)-th most frequently accessed account as:
\(P(k) = \frac{1 / k^{\alpha}}{\sum_{i=1}^{N} 1 / i^{\alpha}}\), which follows a Zipf distribution. 
In the formulation \(N\) denotes the total number of accounts, and \(\alpha\) is the skewness parameter that controls the concentration of the distribution. When \(\alpha = 0\), the access probability follows a uniform distribution. As \(\alpha\) increases, access becomes more concentrated on top-ranked hot accounts, resulting in a more skewed distribution. 

\subsection{Optimization Objective}
Our objective is to determine an optimal parameter configuration \(X^*\) under a given workload \(W\), such that the system performance metric \(J\) is maximized. 
The performance objective function \(J(X; W)\) can be defined according to specific optimization requirements. For example:
\[
\begin{aligned}
J &= \mathrm{throughput}(X; W) && \text{maximize throughput}\\
J &= -\mathrm{latency}(X; W) && \text{minimize average latency}\\
J &= \textstyle\sum_{i=1}^{n} \omega_i f_i(X; W) && \text{multi-objective optimization}
\end{aligned}
\]
The overall optimization problem is therefore formulated as:
\[
X^* = \arg \max_{X \in \Omega} J(X; W).
\]

In the formulation, \(W = (\mu, \beta, \bar{r}, \bar{w}, \alpha)\) denotes the workload feature vector, which characterizes workload scale (average QPS \(\mu\) and variability \(\beta\)), transaction type (average read intensity \(\bar{r}\) and average write intensity \(\bar{w}\)), and access pattern (access skewness \(\alpha\)).
\(X = (x_1, x_2, \cdots, x_n)\) represents a complete system parameter configuration, including tunable parameters across the Execution, Ordering, and Validation phases. \(\Omega\) denotes the feasible configuration space.

\section{EchoFlow Parameter Tuning Framework}

\subsection{Framework}

\vspace{-5mm}
\begin{figure}[htbp]
\centering
\includegraphics[width=\linewidth]{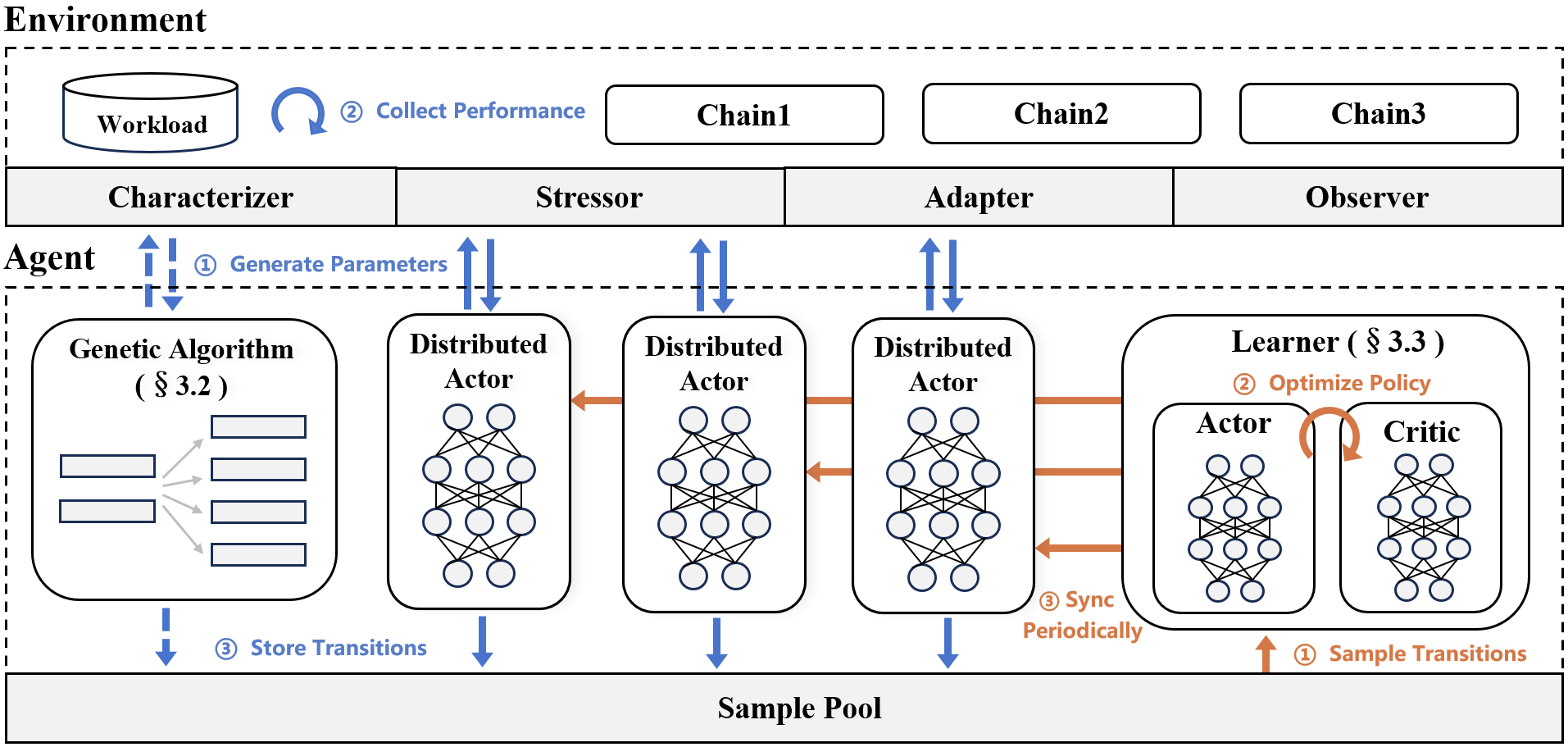}
\caption{The overview of EchoFlow}
\label{fig:fw}
\end{figure}
\vspace{-5mm}
To enable workload-aware automatic parameter tuning, we design and implement the EchoFlow framework.
The overall framework of \textbf{EchoFlow} is illustrated in Fig~\ref{fig:fw}. Similar to most DRL-based frameworks, it consists of two primary components: the \textbf{Agent} and the \textbf{Environment}.

The Environment consists of multiple blockchain instances deployed under identical system conditions. Each blockchain instance can receive and execute transactions under predefined workloads. Through interaction with the Environment, the Agent collects experience samples.

The Agent is further decomposed into multiple \textbf{Distributed-Actors} and a \textbf{Learner}. These components execute asynchronously.  The Learner is the central training component which uses a distributed deterministic policy gradient (D4PG\cite{barth2018d4pg}) approach. It samples experiences from a shared replay buffer, updates the network parameters, and periodically synchronizes the updated Actor parameters to the Distributed-Actors. The Distributed-Actors serve as executors. They interact with the Environment in parallel and continuously store collected experience data into the shared replay buffer.

In addition, we incorporate a genetic algorithm-based parameter tuning approach (GA-Tune) within the Agent. At the initial stage, GA-Tune generates a large number of high-quality parameter configurations and interacts with the environment. Once the GA-Tune no longer produces performance improvements, subsequent parameter configurations are generated by the Distributed-Actors.

Within the Agent--Environment interaction framework, several auxiliary components ensure stable and effective system operation.
The Adapter converts the action vector into parameter configurations and applies them by updating system settings and restarting relevant blockchain services.
The Stressor generates transactions according to predefined workloads to simulate realistic operational pressure.
The Observer monitors system performance, collects key metrics, and computes the corresponding reward.
The Characterizer extracts workload features to form the state representation.

\subsection{Genetic Algorithm-based Tuning}

During the early training stage, the Learner is not yet sufficiently trained, causing the Distributed-Actors to generate low-quality parameter configurations with performance close to random selection. Interacting with the environment using these parameter configurations reduces training efficiency. To address this problem, we introduce \textbf{GA-Tune}, a genetic algorithm-based parameter tuning method. GA-Tune continuously generates superior offspring configurations through population evolution, thereby rapidly producing a set of high-quality parameter configurations and  accelerating the initial training phase of the Learner.

Among various heuristic algorithms, we select the genetic algorithm due to the following advantages: (1) As a widely adopted method, it simulates natural selection and genetic inheritance mechanisms, and can achieve fast convergence. (2) It can naturally incorporate expert-recommended configurations by including them in the initial population.

\vspace{-2mm}
\begin{figure}[htbp]
\centering
\includegraphics[width=\linewidth]{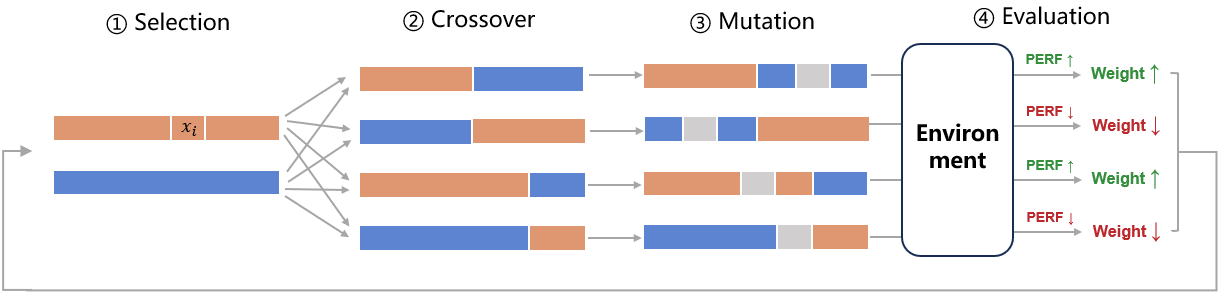}
\caption{the workflow of GA-Tune algorithm}
\label{fig:ga}
\end{figure}
\vspace{-3mm}

The workflow of the GA-Tune algorithm is illustrated in Fig.~\ref{fig:ga}. The algorithm performs initialization by constructing an initial population \(P = \{X_1, \cdots, X_n\}\), where each individual \(X_i\) represents a complete parameter configuration. These configurations can be derived from expert recommendations. 

Subsequently, the algorithm initiates an iterative optimization process starting from the initial population. In each iteration, the algorithm performs the following steps: selecting parent individuals from the current population based on fitness; generating offspring through crossover and mutation; evaluating the offspring and updating their fitness; and merging the offspring with the current population to form an updated population. 

Fitness evaluation serves as the key metric for measuring individual quality. We design a fitness evaluation strategy where the fitness of an individual \(X_i=(x_{i1},x_{i2},\dots,x_{in})\) is defined as the weighted sum of its configuration parameters: \(f(X_i)=\sum_{j=1}^{n} W_{x_{ij}}\). For the initial population, the parameter weights of individual \(X_i\) are determined according to the observed performance of the configuration, denoted as \(\text{PERF}(X_i)\).

During the evolutionary process, the weight update rule for offspring individuals is defined as follows. Suppose an offspring individual \(X\) partially inherits parameters from its parent \(\bar{X}\). Parameters that remain unchanged retain their original weights, while the weights of modified parameters are dynamically adjusted according to their impact on performance:
\[
W_{x_{ij}} =
\begin{cases}
W_{\bar{x}_{ij}}, & x_{ij} = \bar{x}_{ij} \\
W_{\bar{x}_{ij}} + \text{PERF}(X_i) - \text{PERF}(\bar{X}_i), & x_{ij} \ne \bar{x}_{ij}.
\end{cases}
\]

\subsection{D4PG-based Dynamic Tuning}

To mitigate the substantial time required for sample generation in blockchain parameter tuning, we propose \textbf{D4PG-Tune}, a D4PG-based dynamic parameter tuning algorithm. D4PG\cite{barth2018d4pg} is an extension of the DDPG that enhances learning efficiency by incorporating distributed actors. Owing to its off-policy nature and parallel data collection via multiple actors, D4PG effectively improves the sample collection efficiency. In the following, we introduce the algorithm.

\textbf{State.}
The state consists of workload characteristics and physical resource utilization. The workload characteristics include workload scale (average QPS \(\mu\) and volatility \(\beta\)), transaction type (average read intensity \(\bar{r}\) and average write intensity \(\bar{w}\)), and access pattern (skewness parameter \(\alpha\)). Physical resource utilization includes CPU and bandwidth usage.

\textbf{Action.}
The action is defined as a new configuration of all tunable parameters. 
Each parameter must remain within a predefined feasible range to prevent system instability caused by invalid configurations.

\textbf{Reward.}
The reward function guides the Agent toward learning parameter configurations that yield superior performance. 
Unlike prior studies\cite{li2023autotuning,lin2024tunechain}, our reward formulation must explicitly account for workload variations, as the upper bound of achievable performance differs substantially across different workload conditions. Therefore, we define the reward as:
\[
r = \exp\left(\frac{J(X;W) - J(X_d;W)}{J(X_d;W)}\right)
\]
where \(J(X;W)\) and \(J(X_d;W)\) denote the performance under the workload \(W\) using the adjusted parameter configuration \(X\) and the default configuration \(X_d\), respectively. By measuring performance improvement relative to the default configuration under identical workload conditions, this design ensures fair reward evaluation across heterogeneous workloads. Furthermore, the exponential function amplifies differences in performance improvement.

\begin{algorithm}[H]
\caption{D4PG-based Parameter Tuning Algorithm}
\label{alg:d4pg_tuning}
\textbf{Require:} batch size $M$, replay size $R$, exploration constant $\epsilon$, maximum episode number $T$

\textbf{Ensure:} Optimized policy parameters $\theta$

\vspace{0.3em}
\hrule
\vspace{0.3em}

\textbf{Distributed-Actor}
\begin{algorithmic}[1]

\STATE Initialize actor network $\theta_d \gets \theta$
\REPEAT
    \STATE Receive current workload $\mathbf{s}$ from Characterizer
    \STATE Generate action
          $\mathbf{a} = \pi_{\theta_d}(\mathbf{s}) + \epsilon \mathcal{N}(0,1)$
    \STATE Adapter applies $\mathbf{a}$, Stressor generates requests, Observer obtains reward $r$
    \STATE Store transition $(\mathbf{s}, \mathbf{a}, r, \mathbf{s'})$ in the replay buffer
\UNTIL{learner terminates}
\end{algorithmic}

\vspace{0.3em}
\hrule
\vspace{0.3em}

\textbf{Learner}
\begin{algorithmic}[1]
\STATE Initialize actor network $\theta$ and critic network $w$
\STATE Set target networks $(\theta', w') \gets (\theta, w)$
\FOR{$t = 1$ \TO $T$ \textbf{and} $\textbf{when} \ |R| \ge M$}
    \STATE Sample $M$ transitions 
           $ (\mathbf{s}_i, \mathbf{a}_i, r_i, \mathbf{s}'_i) $
           from the replay buffer
    \STATE Compute target distributions $Y_i$ 
    \STATE Update critic network $w$ by $\delta_w$ and actor network $\theta$ by $\delta_\theta$
    \STATE Periodically update target networks:
           $(\theta', w') \gets (\theta, w)$ 
    \STATE Periodically update Distributed-Actor network:
           $\theta_d \gets \theta$ 
\ENDFOR
\RETURN optimized policy parameters $\theta$ 
\end{algorithmic}
\end{algorithm}

D4PG-Tune consists of two core components, Distributed-Actors and Learner, which operate asynchronously. The execution processes of the Distributed-Actor and the Learner are summarized in Algorithm~\ref{alg:d4pg_tuning}.

\textbf{Distributed-Actor.}
The Distributed-Actor contains an actor network that does not perform local training. This actor network continuously interacts with the Environment to collect samples. Multiple Distributed-Actors can run in parallel, thereby significantly accelerating sample collection.

First, the Distributed-Actor copies the parameters \(\theta\) from the Learner's actor network as its initial parameters (line 1). It then enters an interaction loop with the Environment to collect training samples (lines 2--7). In each iteration, the Distributed-Actor generates a new parameter configuration (action) based on the current workload (state), and then obtain the performance results (reward) under this action. Finally, the Distributed-Actor stores the transition \((s, a, r, s')\) in the replay buffer for subsequent training by the Learner.

\textbf{Learner.}
The Learner contains an actor and a critic network, each with a corresponding target network. The actor network, parameterized by \(\theta\), represents a deterministic policy \(\pi\) for action generation. 
The critic network, parameterized by \(w\), evaluates the effectiveness of the policy \(\pi\). 

The Learner first initializes the networks and copies their parameters to the target networks (lines 1--2). Once the number of samples in the replay buffer reaches the batch size \(M\), the Learner enters the training loop (lines 3--9). 

In each iteration, the Learner samples \(M\) transitions 
from the replay buffer (line 4). It then computes the target distribution for each sample: \(Y_i = \mathcal{P}\bigl(r_i + \gamma Z_{w'}(\mathbf{s}'_i, \pi_{\theta'}(\mathbf{s}'_i))\bigr)\) (line 5), where \(\mathcal{P}\) denotes the projection operator that maps the distribution onto the predefined support of the critic network 
and \(Z_w\) is a discrete probability distribution that provides more stable value estimates. 

After obtaining the target distribution, the Learner updates the critic network and actor network (line 6). By minimizing the distributional distance \(d\) between the predicted distribution \(Z_w(\mathbf{s}_i, \mathbf{a}_i)\) and the target distribution \(Y_i\), the critic gradient is computed as:
\(
\delta_w = \frac{1}{M} \sum_i (|R| \, p_i)^{-1} \nabla_w d\bigl(Y_i, Z_w(\mathbf{s}_i, \mathbf{a}_i)\bigr).
\)
Then using the Deterministic Policy Gradient method, the actor gradient is computed as:
\(
\delta_\theta =
\frac{1}{M} \sum_i
\left.
\nabla_\theta \pi_\theta(\mathbf{s}_i)
\cdot
\mathbb{E}\big[\nabla_{\mathbf{a}} Z_w(\mathbf{s}_i, \mathbf{a})\big]
\right|_{\mathbf{a} = \pi_\theta(\mathbf{s}_i)}.
\)
In these formulations, \(p_i\) denotes the sampling probability associated with sample \(i\), while \(|R|\) represents the total number of samples currently maintained in the replay buffer.

During training, the Learner periodically coppies the actor and critic parameters to target networks to maintain stability in the target distribution (line 7). Meanwhile, the actor network parameters are also periodically synchronized across all Distributed-Actors, enabling them to interact with the Environment using the latest policy (line 8).

\section{Evaluation}

Our experimental study is designed to address the following Research Questions:

\begin{enumerate}
    \item How much do workload variations affect performance?
    \item To what extent can our method sustain performance improvements under diverse workloads?
    \item How much do D4PG-Tune and GA-Tune accelerate training?
\end{enumerate}

\subsubsection{Experimental Environment.}

Our experiments were conducted on five physical servers equipped with an Intel Xeon Platinum 8255C CPU @2.50\,GHz, 128\,GB RAM, running CentOS 7. We selected Hyperledger Fabric as the experimental platform because it is a widely adopted permissioned blockchain system. A local Fabric network was deployed to evaluate our approach. Each Fabric instance runs Hyperledger Fabric v2.5.14 and consists of two ordering nodes and four peer nodes, with the inter-node bandwidth constrained to 100\,Mb/s. We used Python3 and PyTorch to implement all of our algorithms and components.

\subsubsection{Workload Configuration.}

We deployed the SmallBank smart contract, a benchmark widely used in academic research\cite{sharma2019blurring,dinh2017blockbench}. This contract simulates basic bank-account operations, which facilitates precise control of the read--write ratio. To investigate the impact of workload diversity on system performance, we designed multiple representative workload patterns: read-heavy\textbf{ (RH)} with an average read intensity $\bar{r}=0.8$; write-heavy\textbf{ (WH)} with an average intensity ratio $\bar{w}=0.8$; write-heavy with concentrated account access\textbf{ (WH-C)}, where $\bar{w}=0.8$ and the skewness parameter $\alpha=1$; and read-heavy with query-rate variation\textbf{ (RH-V)}, where $\bar{r}=0.8$ and the volatility $\beta=1$.

\subsection{Effectiveness of Workload Awareness}

\begin{figure}[htbp]
\centering
\includegraphics[width=\linewidth]{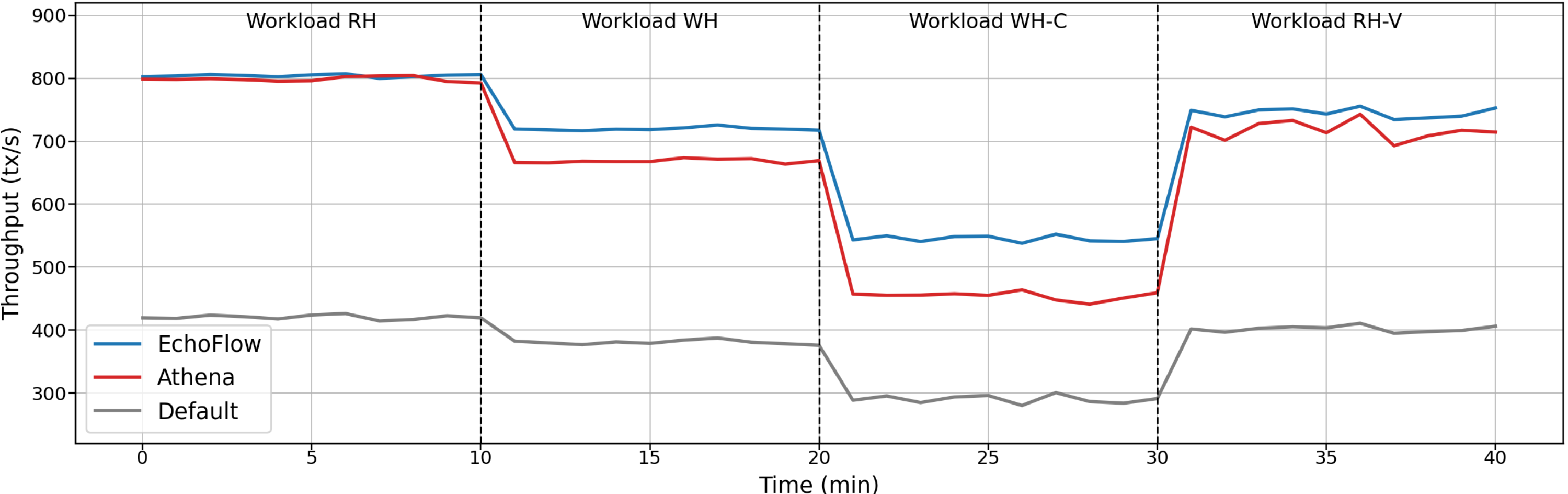}
\caption{Throughput comparison under diverse workloads}
\label{fig:e1}
\end{figure}
\vspace{-5mm}

To address Research Questions~1 and~2, we conducted experiments under diverse workloads, with the results illustrated in Fig.~\ref{fig:e1}.
We evaluated the throughput under three configurations: the default configuration, the configuration generated by Athena, and the dynamic configuration produced by EchoFlow. Athena\cite{li2023autotuning}, a multi-agent DRL method designed to optimize the parameters of different nodes in Fabric, was selected as our primary baseline for comparison. 

Under the default parameter configuration, the average throughput values across the four workloads were 420.50, 380.54, 290.24, and 401.93~tps, respectively, exhibiting substantial performance disparities. Notably, compared with the RH workload, the average throughput under the WH, WH-C, and RH-V workloads decreased by 9.50\%, 30.98\%, and 4.42\%, demonstrating that workload variations have a significant impact on throughput (RQ~1).

We further observed that the parameter configuration provided by Athena significantly improved average throughput. Under the RH workload, Athena achieved 798.48~tps, an 89.89\% improvement over the default configuration. However, as the workload changed, throughput declined to 668.65, 454.46, and 717.49~tps under the WH, WH-C, and RH-V workloads, representing drops of 16.26\%, 43.08\%, and 10.14\%.

In contrast, EchoFlow dynamically adjusted parameters according to workload characteristics. It achieved average throughput values of 804.22, 719.59, 544.98, and 745.24~tps under diverse workloads. Compared with Athena, EchoFlow improved throughput by 7.62\%, 19.92\%, and 3.87\% as the workload changed, demonstrating better adaptability and sustained performance gains (RQ~2).

\subsection{Acceleration of Training}

\begin{figure}[H]
\centering
\includegraphics[width=0.7\linewidth]{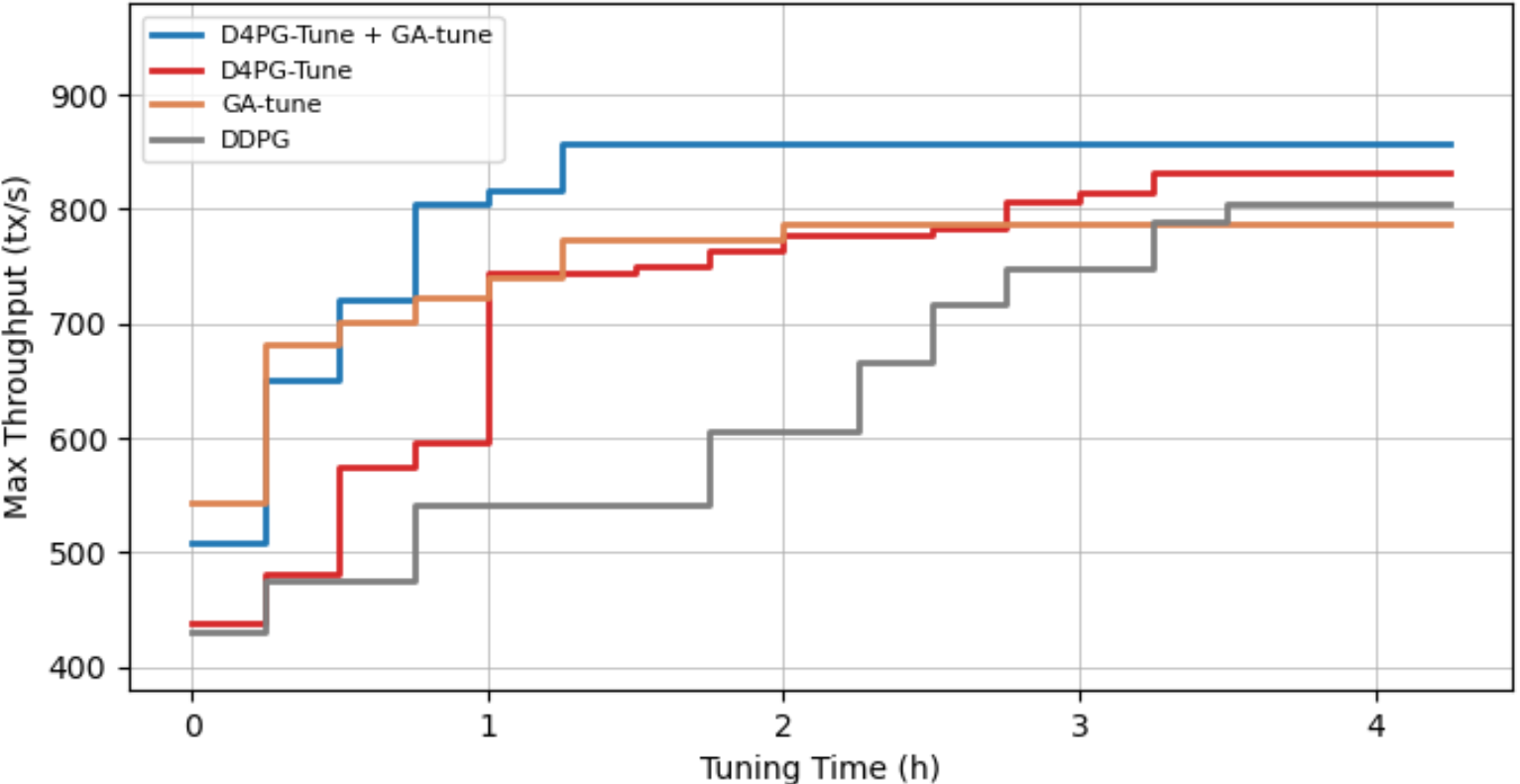}
\caption{Best throughput achieved over time by different methods}
\label{fig:e2}
\end{figure}
\vspace{-5mm}

To answer Research Question~3, we conducted experiments under a fixed workload to evaluate training efficiency: (i) the proposed method (D4PG-Tune + GA-Tune), (ii) D4PG-Tune only, (iii) GA-Tune only, and (iv) the baseline DDPG without enhancement. During training, we recorded the best throughput achieved so far every 15 minutes, as illustrated in Fig.~\ref{fig:e2}.

When using the baseline DDPG method, the model required about 3.5 hours to converge, and exhibited limited early-stage improvement due to insufficient exploration samples. With D4PG-Tune, multiple samples could be collected in parallel, accelerating the training process. After 1 hour of training, the best throughput reached 745.02~tps. Under the same training duration, D4PG-Tune substantially outperformed the baseline. Nevertheless, it still required approximately 3 hours to converge to its optimal performance.

To further accelerate optimization, we introduced GA-Tune. By incorporating expert configurations into the initial population, GA-Tune achieved strong initial performance, reaching 701.38~tps within only 0.5 hours. 
However, GA-Tune alone tends to converge prematurely to local optima, reaching only 786.93~tps. To address this, our proposed method integrates GA-Tune with D4PG-Tune, enabling convergence within 1.25 hours and achieving 857.60~tps, demonstrating faster training and superior parameter configurations (RQ~3).

\section{Related Work}

Automatic parameter tuning has been widely studied in the database domain and can generally be categorized into three approaches: search-based\cite{ansel2014opentuner,zhu2017bestconfig}, model-based\cite{vanaken2017ottertune,bao2018autoconfig}, and DRL-based methods\cite{zhang2019cdbtune,li2019qtune,cai2022hunter}. Search-based methods rely on heuristic search strategies to efficiently identify high-performance configurations, though they may suffer from local optima. Model-based methods build predictive models to capture the relationship between configuration parameters and system performance, but they require large volumes of high-quality training data. In contrast, DRL-based methods interact with the environment through trial-and-error learning, requiring fewer samples and discovering improved configurations, making them a promising research direction. 

Inspired by recent advances in database autotuning, blockchain parameter tuning has increasingly adopted DRL-based techniques due to the high cost of collecting sufficient training data. Athena\cite{li2023autotuning} is the first to apply a DRL-based method to optimize the parameters of different node types in Hyperledger Fabric, significantly improving throughput and reducing latency. TuneChain\cite{lin2024tunechain} enables online tuning by training multiple models to adapt to different transaction conflict ratios. 
In addition, several DRL-based approaches go beyond parameter tuning and focus on broader blockchain optimization tasks. SPRING\cite{wu2024spring} optimizes the shard placement of states, aiming to minimize the proportion of cross-shard transactions while maintaining workload balance among shards. AdaChain\cite{wu2023adachain} adapts blockchain system architectures and key parameters, enabling the system to better match deployed smart contracts and workload characteristics.

\section{Conclusion}
In this paper, we propose EchoFlow, a workload-aware dynamic parameter tuning approach for blockchain systems. EchoFlow adaptively adjusts system configurations according to workload characteristics, enabling continuous performance optimization. To mitigate the substantial time required for sample generation in blockchain environments, we adopt a distributed reinforcement learning method that integrates parallel data collection from multiple actors, substantially improving training efficiency. Furthermore, a genetic algorithm is introduced during the initial training stage to generate high-quality samples, shortening the early exploration phase. Extensive experimental results demonstrate that, compared with existing approaches, EchoFlow consistently achieves more stable and superior performance across diverse workloads while also reducing training time.

\vspace{0.7em}
\noindent \textbf{Acknowledgements.} This work was supported in part by the National Natural Science Foundation of China under Grant U22B2032.

\bibliographystyle{spmpsci_unsrt}
\bibliography{reference}

\begin{thebibliography}{10}
\providecommand{\url}[1]{{#1}}
\providecommand{\urlprefix}{URL }
\expandafter\ifx\csname urlstyle\endcsname\relax
  \providecommand{\doi}[1]{DOI~\discretionary{}{}{}#1}\else
  \providecommand{\doi}{DOI~\discretionary{}{}{}\begingroup \urlstyle{rm}\Url}\fi

\bibitem{hyperledgerFabricDocs}
{Hyperledger Foundation}: {Hyperledger Fabric Documentation} (2026).
\newblock \urlprefix\url{https://hyperledger-fabric.readthedocs.io/en/release-2.5/}.
\newblock Accessed: 2026-03-04

\bibitem{sharma2019blurring}
Sharma, A., Schuhknecht, F.M., Agrawal, D., Dittrich, J.: Blurring the lines between blockchains and database systems: the case of hyperledger fabric.
\newblock In: Proceedings of the 2019 International Conference on Management of Data, pp. 105--122 (2019)

\bibitem{xu2023adaptchain}
Xu, J., Xie, Q., Peng, S., Wang, C., Jia, X.: Adaptchain: adaptive scaling blockchain with transaction deduplication.
\newblock IEEE Transactions on Parallel and Distributed Systems \textbf{34}(6), 1909--1922 (2023)

\bibitem{li2023autotuning}
Li, M., Wang, Y., Ma, S., Liu, C., Huo, D., Wang, Y., Xu, Z.: Auto-tuning with reinforcement learning for permissioned blockchain systems.
\newblock Proceedings of the VLDB Endowment \textbf{16}(5), 1000--1012 (2023)

\bibitem{lin2024tunechain}
Lin, J., Deng, R., Lu, Z., Zhang, Y., Duan, Q.: {TuneChain}: an online configuration auto-tuning approach for permissioned blockchain systems.
\newblock In: Proceedings of the 2024 IEEE International Conference on Web Services (ICWS), pp. 512--523 (2024)

\bibitem{zhang2024morphdag}
Zhang, S., Xiao, J., Wu, E., Cheng, F., Li, B., Wang, W., Jin, H.: {MorphDAG}: a workload-aware elastic dag-based blockchain.
\newblock IEEE Transactions on Knowledge and Data Engineering \textbf{36}(10), 5249--5264 (2024)

\bibitem{khan2025scalability}
Khan, M.M., Khan, F.S., Nadeem, M., Khan, T.H., Haider, S., Daas, D.: Scalability and efficiency analysis of hyperledger fabric and private ethereum in smart contract execution.
\newblock Computers \textbf{14}(4), 132 (2025)

\bibitem{barth2018d4pg}
Barth-Maron, G., Hoffman, M.W., Budden, D., Dabney, W., Horgan, D., TB, D., Muldal, A., Heess, N., Lillicrap, T.: Distributed distributional deterministic policy gradients.
\newblock In: International Conference on Learning Representations (ICLR) (2018)

\bibitem{dinh2017blockbench}
Dinh, T.T.A., Wang, J., Chen, G., Liu, R., Ooi, B.C., Tan, K.L.: {BLOCKBENCH}: a framework for analyzing private blockchains.
\newblock In: Proceedings of the 2017 ACM International Conference on Management of Data, pp. 1085--1100 (2017)

\bibitem{ansel2014opentuner}
Ansel, J., Kamil, S., Veeramachaneni, K., Ragan-Kelley, J., Bosboom, J., O'Reilly, U.M., Amarasinghe, S.: {OpenTuner}: an extensible framework for program autotuning.
\newblock In: Proceedings of the 23rd International Conference on Parallel Architectures and Compilation Techniques (PACT), pp. 303--316 (2014)

\bibitem{zhu2017bestconfig}
Zhu, Y., Liu, J., Guo, M., Bao, Y., Ma, W., Liu, Z., Song, K., Yang, Y.: {BestConfig}: tapping the performance potential of systems via automatic configuration tuning.
\newblock In: Proceedings of the 2017 ACM Symposium on Cloud Computing, pp. 338--350 (2017)

\bibitem{vanaken2017ottertune}
Van~Aken, D., Pavlo, A., Gordon, G.J., Zhang, B.: Automatic database management system tuning through large-scale machine learning.
\newblock In: Proceedings of the 2017 ACM International Conference on Management of Data, pp. 1009--1024 (2017)

\bibitem{bao2018autoconfig}
Bao, L., Liu, X., Xu, Z., Fang, B.: Autoconfig: automatic configuration tuning for distributed message systems.
\newblock In: Proceedings of the 33rd ACM/IEEE International Conference on Automated Software Engineering, pp. 29--40 (2018)

\bibitem{zhang2019cdbtune}
Zhang, J., Liu, Y., Zhou, K., Li, G., Xiao, Z., Cheng, B., Xing, J., Wang, Y., Chen, T., Liu, L., Ran, M., Li, Z.: An end-to-end automatic cloud database tuning system using deep reinforcement learning.
\newblock In: Proceedings of the 2019 International Conference on Management of Data, pp. 415--432 (2019)

\bibitem{li2019qtune}
Li, G., Zhou, X., Li, S., Gao, B.: Qtune: a query-aware database tuning system with deep reinforcement learning.
\newblock Proceedings of the VLDB Endowment \textbf{12}(12), 2118--2130 (2019)

\bibitem{cai2022hunter}
Cai, B., Liu, Y., Zhang, C., Zhang, G., Zhou, K., Liu, L., Li, C., Cheng, B., Yang, J., Xing, J.: Hunter: an online cloud database hybrid tuning system for personalized requirements.
\newblock In: Proceedings of the 2022 International Conference on Management of Data, pp. 646--659 (2022)

\bibitem{wu2024spring}
Li, P., Song, M., Xing, M., Xiao, Z., Ding, Q., Guan, S., Long, J.: Spring: improving the throughput of sharding blockchain via deep reinforcement learning based state placement.
\newblock In: Proceedings of the ACM Web Conference 2024, pp. 2836--2846 (2024)

\bibitem{wu2023adachain}
Wu, C., Mehta, B., Amiri, M.J., Marcus, R., Loo, B.T.: Adachain: a learned adaptive blockchain framework.
\newblock Proceedings of the VLDB Endowment \textbf{16}(8), 2033--2046 (2023)

\end{thebibliography}

\end{document}